\documentclass[11pt]{article}

\usepackage{graphicx,amsmath,amsfonts,amssymb,dcolumn,amsthm,slashed,bbm,xspace,pgffor,tikz,mathbbol,latexsym,enumerate}
\usepackage{soul}
\usepackage[colorlinks,citecolor=red,urlcolor=cyan,linkcolor=blue]{hyperref}
\usepackage[utf8]{inputenc}
\usepackage[english]{babel}
\usepackage{lmodern}
\usepackage{microtype}
\usepackage{cite}

\numberwithin{equation}{section}

\begin{document}

\title{\textbf{Non-relativistic limit of gravity theories in the first order formalism}}

\author{\textbf{Amanda Guerrieri}\thanks{amguerrieri@id.uff.br} \ and \textbf{Rodrigo F.~Sobreiro}\thanks{rodrigo\_sobreiro@id.uff.br}\\\\
\textit{{\small UFF - Universidade Federal Fluminense, Instituto de F\'isica,}}\\
\textit{{\small Av. Litorânea s/n, 24210-346, Niter\'oi, RJ, Brasil.}}}

\date{}
\maketitle

\begin{abstract}
We consider the non-relativistic limit of gravity in four dimensions in the first order formalism. First, we revisit the case of the Einstein-Hilbert action and formally discuss some geometrical configurations in vacuum and in the presence of matter at leading order. Second, we consider the more general Mardones-Zanelli action and its non-relativistic limit. The field equations and some interesting geometries, in vacuum and in the presence of matter, are formally obtained. Remarkably, in contrast to the Einstein-Hilbert limit, the set of field equations is fully determined because the boost connection appears in the action and field equations. It is found that the cosmological constant must disappear in the non-relativistic Mardones-Zanelli action at leading order. The conditions for Newtonian absolute time be acceptable are also discussed. It turns out that Newtonian absolute time can be safely implemented with reasonable conditions.
\end{abstract}

\newpage

\tableofcontents

\newpage

\section{Introduction}

One of the most important features of a new larger theory is that it must be reduced to the known, well established, smaller theory at some particular limit. General Relativity (GR) is a celebrated example \cite{Misner:1974qy,Wald:1984rg,DeSabbata:1986sv}. In fact, Newtonian gravity can be obtained from GR by a suitable weak field limit. In the context of GR described as a gauge theory for the Poincar\'e group\footnote{More precisely, only the Lorentz sector is gauged, since the Poincar\'e group is not a semi-simple Lie group. Nevertheless, the translational sector is used to expand the vierbein degrees of freedom, which is fundamental to encode the equivalence principle \cite{Utiyama:1956sy,Kibble:1961ba,Sciama:1964wt,Mardones:1990qc,Zanelli:2005sa}.}  \cite{Utiyama:1956sy,Kibble:1961ba,Sciama:1964wt}, the so-called first order formalism, GR can be reduced to Newtonian gravity by a appropriate In\"on\"u-Wigner (IW) contraction \cite{Inonu:1953sp}, see for instance \cite{Bergshoeff:2017btm,Banerjee:2018gqz} and references therein. On the one hand, one can take the ultra-relativistic limit of the Poincaré group and end up in the so-called Carrol gravity. On the other hand, the non-relativistic limit of the Poincar\'e group leads to the Galilei group and Galilei gravity. Galilei gravity is then a gauge theory based on the Galilei group and describes a geometrodynamical theory which, under appropriate assumptions \cite{Christensen:2013rfa,Banerjee:2014nja,Afshar:2015aku,Abedini:2019voz}, is equivalent to Newton theory of gravity. An important condition is known as \emph{twistless torsion constraint} which is necessary to define the Newtonian absolute time. Nevertheless, Galilei gravity is far more general than Newtonian gravity, carrying, for instance, torsion degrees of freedom \cite{Banerjee:2016laq,Bergshoeff:2017dqq}. For the earlier works in NC geometrodynamics we refer to the seminal papers of E.~Cartan \cite{Cartan:1923zea,Cartan:1924yea} and also \cite{Trautman:1963aaa,Havas:1964zza,Trautman:1965aaa,Kunzle:1972aaa,Misner:1974qy,Dixon:1975fy}. In contrast to the non-relativistic limit of GR as the gauge theory for the Galilei group, one can start directly from the Galilei algebra and its extensions in order to construct more general Newton-Cartan (NC) gravity theories \cite{Duval:1984cj,Duval:2009vt,Bergshoeff:2019ctr}. Two renowned examples are the Bargmann and Shr\"odinger groups. A particularly appealing feature of these extensions \cite{Christensen:2013rfa,Afshar:2015aku,Hartong:2015zia} is that it can be used as a base for constructing Ho$\check{\mathrm{r}}$ava-Lifshitz gravity models \cite{Horava:2009uw}. Moreover, NC geometries have renowned importance in string and brane scenarios, see for instance \cite{Roychowdhury:2019sfo,Kluson:2020aoq,Bergshoeff:2019pij,Bergshoeff:2020oes} and references therein.

The gauge approach to describe GR allows torsion degrees of freedom to enter in the context in a natural way \cite{Utiyama:1956sy,Kibble:1961ba,Sciama:1964wt,Mardones:1990qc,Zanelli:2005sa}. Moreover, the differential form language shows itself to be a useful and natural framework in this scenario because it allows to work in a coordinate independent manner. This approach enables one to write down more general actions, going beyond the Einstein-Hilbert (EH). In fact, in \cite{Mardones:1990qc} Mardones and Zanelli were able to generalize Lovelock gravity theory \cite{Lovelock:1971yv} in order to account for torsion terms in any spacetime dimension. These theories, known as Lovelock-Cartan (LC) gravities, are described by first order, and polynomially local actions with explicit Lorentz gauge symmetry, called Mardones-Zanelli (MZ) actions\footnote{In this work we equally refer to such theories as MZ or LC theories.}. Moreover, in differential form language, MZ actions do not depend explicitly on the metric tensor. The aim of the present work is to develop the non-relativistic limit of such theories and go beyond EH action in the study of non-relativistic gravity systems. In this first endeavor, we stick to four dimensions. In four-dimensional MZ action, besides the EH term, gravity gains topological terms (which do not affect the classical dynamics of gravity), a cosmological constant term and two extra terms associated with torsion. Thus, the present study opens the door in the understanding of the possible effects of the cosmological constant and torsion (or if the lack of them can be accommodated in the non-relativistic MZ theory) in non-relativistic gravitational systems. 

In LC gravities, the EH action is a particular case. Hence, before considering the four-dimensional LC theory we review the EH case. Galilei gravity is obtained from the EH action by employing the appropriate IW contraction of the Poincaré group. The field equations are derived and solutions are formally discussed. We analyze torsionless and torsional solutions in vacuum and in the presence of matter in a quite general way. Particularly, we discuss twistless torsion conditions in the presence of other torsion components. Furthermore, an explicit example beyond Newtonian absolute time is worked out. In this example, we find a Weintzenb\"ock-like geometry \cite{Weitzenbock:1923boo,Garecki:2010jj,Aldrovandi:2013wha,Golovnev:2018red} with conformally flat metrics.

As mentioned before, the main target of the present work is the four-dimensional full MZ action. Its non-relativistic limit is performed and the field equations derived. Obviously, the residual gauge symmetry is again described by the Galilei group. To distinguish from the Galilei gravity obtained from the EH action, we call the non-relativistic theory obtained from the MZ action by \emph{Galilei-Cartan} (GC) gravity. The first interesting result is that, differently from the EH case, the boost connection is explicitly present at the action and field equations. Thence, the system of field equations are fully determined. Another result is the establishment of the properties that the GC theory must fulfill in order to accept Newtonian absolute time. Solutions in vacuum and in the presence of matter, with and without torsion, are formally discussed. Particularly, the torsionless vacuum solution can be adjusted in order to account for the non-relativistic limit of the maximally symmetric solution of the MZ relativistic theory. It is also shown (In Appendix \ref{CCap}) that if we keep the cosmological constant in the GC theory, the field equations are inconsistent with the Bianchi identities. Thus, the cosmological constant must indeed vanish in the non-relativistic limit. This result ensures that the final form of the GC action with no cosmological constant is the only possibility at leading order.

This work is organized as follows: In Section \ref{LC} we review the construction of the LC theory of gravity in four dimensions and the maximally symmetric vacuum solution is derived. In Section \ref{IW} a discussion about the IW contraction from the Poincaré group towards Galilei group is provided. As an intermediate illustrative step, the four-dimensional non-relativistic limit of the EH action is obtained in Section \ref{EH}. Also in this section, some novel solutions are discussed. The non-relativistic limit of the four-dimensional LC theory of gravity is considered in Section \ref{GC}. Formal solutions in vacuum and in the presence of matter distributions are obtained. the acceptance of Newtonian absolute time are also analyzed in this section.  Our conclusions are displayed in Section \ref{FINAL}. Moreover, the fate of the cosmological constant is discussed in the Appendix \ref{CCap}.

\section{Lovelock-Cartan gravity}\label{LC}

In this section we review the construction of the LC theory of gravity in four dimensions \cite{Mardones:1990qc,Zanelli:2005sa} which generalizes Lovelock gravity theories \cite{Lovelock:1971yv} within Einstein-Cartan first order formalism by including torsional terms in the gravity action. The scenario is a four-dimensional manifold $M$ with local Minkowski metric, $\eta=\mathrm{diag}(-,+,+,+)$. The construction is fundamentally based on the 10 parameter Poincaré group $ISO(1,3)=SO(1,3)\times\mathbb{R}^{1,3}$, whose algebra is given by
\begin{eqnarray}
\left[\Sigma_{AB},\Sigma_{CD}\right]&=&\frac{1}{2}\left(\eta_{AD}\Sigma_{BC}-\eta_{AC}\Sigma_{BD}+\eta_{BC}\Sigma_{AD}-\eta_{BD}\Sigma_{AC}\right)\;,\nonumber\\
\left[\Sigma_{AB},P_C\right]&=&\frac{1}{2}\left(\eta_{BC}P_A-\eta_{AC}P_B\right)\;,\nonumber\\
\left[P_A,P_B\right]&=&0\;,\label{Poinc1}
\end{eqnarray}
with $\Sigma_{AB}=-\Sigma_{BA}$ being the 6 generators of the Lorentz group $SO(1,3)$ and $P_A$ the 4 generators of the translational group $\mathbb{R}^{1,3}$. Capital Latin indices run through $\{0,1,2,3\}$. The Poincaré group is not a semi-simple Lie group, a property that makes the task of constructing a gauge invariant action quite difficult\footnote{Some exceptions do exist. For instance, in odd dimensions it is possible to work with Chern-Simons theories, see \cite{Mardones:1990qc}.}. A solution for this problem is to consider only the Lorentz sector for gauging while keeping the translational sector of the group to expand a vector representation of the Lorentz gauge group. Aiming the construction of a general gravity action, we consider the fundamental independent 1-form fields: the vierbein $E$ and the Lorentz connection $Y$, defined respectively through
\begin{eqnarray}
E&=&E^AP_A\;,\nonumber\\
Y&=&Y^{AB}\Sigma_{AB}\;.\label{4fields1}
\end{eqnarray}
The vierbein $E^A$ ensures the equivalence principle by defining a local isomorphism between generic spacetime coordinates $x^\mu$ (Lowercase Greek indices refer to spacetime indices and run through $\mu,\nu,\alpha\ldots\in\{0,1,2,3\}$) with inertial coordinates $x^A$. The latter can be identified with coordinates in the tangent space $T(M)$ at the point $x\in M$. The Lorentz connection $Y^{AB}$ carries information about the parallel transport on $M$ because it is directly related to the affine connection on $M$. Since $E$ and $Y$ are independent fields, the geometric properties of metricity and parallelism are also independent concepts in the first order formalism. The corresponding 2-form field strengths constructed out from $E$ and $Y$ are the curvature and torsion, respectively defined by\footnote{A comment about the notation is important at this point. We extensively employ differential forms framework. Thus, the wedge operator is assumed in most expressions of the paper (except when tensor notation is used). For example, expression \eqref{4fields2} must be read as $\Omega^{AB}=d\wedge Y^{AB}+Y^A_{\phantom{A}C}\wedge Y^{CB}$. In expression \eqref{inv0} however no wedge product is assumed.}
\begin{eqnarray}
\Omega^{AB}&=&dY^{AB}+Y^A_{\phantom{A}C}Y^{CB}\;,\nonumber\\
W^A&=&\nabla E^A\;\;=\;\;dE^A+Y^A_{\phantom{A}B}E^B\;,\label{4fields2}
\end{eqnarray}
with $\nabla\cdot\equiv d\cdot+[Y,\cdot]$ standing for the Lorentz covariant derivative. The Bianchi hierarchy relations are easily found,
\begin{eqnarray}
W^A&=&\nabla E^A\;,\nonumber\\
\nabla W^A&=&\Omega^A_{\phantom{A}B} E^B\;,\nonumber\\
\nabla\Omega^A_{\phantom{A}B}&=&0
\;.\label{hier1}
\end{eqnarray}
The gauge (local Lorentz) transformations are defined through
\begin{eqnarray}
Y^\prime&=&u^{-1}(d+Y)u\;\;\approx\;\;Y+\nabla\alpha\;,\nonumber\\
E^\prime&=&u^{-1}Eu\;\;\approx\;\;E+[E,\alpha]\;,\label{gt0}
\end{eqnarray}
where $u=\exp\alpha\approx1+\alpha$ is a Lorentz group element and $\alpha=\alpha^{AB}\Sigma_{AB}$ is assumed to be an infinitesimal algebra-valued parameter\footnote{The gauge parameter does not need to be infinitesimal, MZ actions are generally invariant under finite gauge transformations.}.

The inverse of the vierbein $E_A$ is assumed to exist in such a way that
\begin{eqnarray}
E^A_\mu E^\mu_B&=&\delta^A_B\;,\nonumber\\
E^A_\mu E_A^\nu&=&\delta_\mu^\nu\;.\label{inv0}
\end{eqnarray}
Moreover, spacetime and locally invariant metrics, $g_{\mu\nu}$ and $\eta^{AB}$, respectively, obey
\begin{eqnarray}
\eta^{AB}&=&E^A_\mu E^B_\nu g^{\mu\nu}\;,\nonumber\\
\eta_{AB}&=&E_A^\mu E_B^\nu g_{\mu\nu}\;,\nonumber\\
g_{\mu\nu}&=&E^A_\mu E^B_\nu\eta_{AB}\;,\nonumber\\
g^{\mu\nu}&=&E_A^\mu E_B^\nu\eta^{AB}\;,\label{metrics0}
\end{eqnarray}
with $g_{\mu\alpha}g^{\alpha\nu}=\delta^\nu_\mu$ and $\eta_{AC}\eta^{CB}=\delta^B_A$. Moreover, besides $\delta_A^B$, we have the gauge invariant skew-symmetric object $\epsilon_{ABCD}$, the Levi-Civita symbol at our disposal.

The MZ theorem \cite{Mardones:1990qc,Zanelli:2005sa} states that the most general four-dimensional action which is gauge invariant, polynomially local, explicitly metric independent, and that depends only on first order derivatives is given by
\begin{eqnarray}
    S_{MZ}&=&\kappa\int\;\epsilon_{ABCD}\left(\Omega^{AB}E^CE^D+\frac{\Lambda}{2}E^AE^BE^CE^D\right)+\int\left(z_1\Omega^{AB}E_AE_B+z_2W^AW_A\right)+\nonumber\\
    &+&\int\left(z_3\epsilon_{ABCD}\Omega^{AB}\Omega^{CD}+z_4\Omega^{AB}\Omega_{AB}\right)+S_m\;.\label{mzaction1}
\end{eqnarray}
The first term is the usual EH action while the second is the cosmological constant term. Hence, $\kappa$ is related to the inverse of Newton's constant and $\Lambda$ is the cosmological constant. Third and fourth terms are essentially the same, up to surface terms. The parameters $z_1$ and $z_2$ carry mass squared dimension and when $z_2=-z_1$ these terms are reduced to the Nieh-Yan topological term \cite{Mardones:1990qc,Zanelli:2005sa,Nieh:1981ww,Nieh:2007zz,Nieh:2018rlg}. The fifth and six terms are, respectively, Gauss-Bonnet and Pontryagin topological terms \cite{Mardones:1990qc,Zanelli:2005sa,Kobayashi,Nakahara:1990th}. Being topological, these terms do not contribute to the field equations. Clearly, $z_3$ and $z_4$ are dimensionless topological parameters. Finally, $S_m$ stands for the matter content of fields and particles which, up to some considerations, we will keep as general as possible in this paper. 

The field equations are quite easily obtained by varying the MZ action \eqref{mzaction1} with respect to $E^A$ and $Y^{AB}$, respectively,
\begin{eqnarray}
\kappa\epsilon_{ABCD}\left(\Omega^{BC}E^D+\Lambda E^BE^CE^D\right)+(z_1+z_2)\Omega_{AB}E^B&=&-\frac{1}{2}\frac{\delta S_m}{\delta E^A}\;,\nonumber\\
\kappa\epsilon_{ABCD}W^CE^D+\frac{(z_1+z_2)}{2}\left(W_AE_B-E_AW_B\right)&=&-\frac{1}{2}\frac{\delta S_m}{\delta Y^{AB}}\;.\label{mzfeq0}
\end{eqnarray}
For further use, we can derive a torsionless vacuum solution of the field equations \eqref{mzfeq0},
\begin{eqnarray}
\Omega^{AB}_0&=&a_0\Lambda E^AE^B-\frac{\kappa}{(z_1+z_2)}(1+a_0)\Lambda\epsilon^{AB}_{\phantom{AB}CD}E^CE^D\;,\nonumber\\
W_0^A&=&0\;,\label{dS00}
\end{eqnarray}
with $a_0\in\mathbb{R}$. It can be easily checked that the substitution of solutions \eqref{dS0} in Bianchi identities \eqref{hier1} leads to $a_0=-1$. Hence, the torsionless vacuum solution \eqref{dS00} of the MZ action is a maximally symmetric spacetime
\begin{eqnarray}
\Omega^{AB}_0&=&-\Lambda E^AE^B\;,\nonumber\\
W_0^A&=&0\;.\label{dS0}
\end{eqnarray}
For a solution with general $a_0$, one should, perhaps, consider non-trivial torsion.

The specific case of GR can be obtained as a particular case of the MZ action \eqref{mzaction1} by setting $\Lambda=z_i=0$, namely
\begin{equation}
    S_{EH}=\kappa\int\epsilon_{ABCD}\Omega^{AB}E^CE^D\;.\label{ehaction0}
\end{equation}
Before we discuss the non-relativistic limit of MZ action in Section \ref{GC}, we will revisit the non-relativistic limit of the Einstein-Hilbert  \eqref{ehaction0} in Section \ref{EH}. But first, let us take a look at the In\"on\"u-Wigner contraction \cite{Inonu:1953sp} from the Poincar\'e group to the Galilei group in the next Section.

\section{In\"on\"u-Wigner contraction of the Poincar\'e group}\label{IW}

The first step towards the non-relativistic limit of actions \eqref{mzaction1} and \eqref{ehaction0} is to split the Poincar\'e group as $ISO(1,3)=SO(3)\times L(3)\times\mathbb{R}_s^3\times\mathbb{R}_t$ where $SO(3)$ describes spatial rotations, $L(3)$ refers to Lorentz boosts, $\mathbb{R}_s^3$ to space translations, and $\mathbb{R}_t$ to time translations. The corresponding algebra is easily obtained by projecting the algebra \eqref{Poinc1} in spatial and temporal components. We define
\begin{eqnarray}
\Sigma_{AB}&\equiv&\left(\Sigma_{ab},\Sigma_{a0}\right)\;\;=\;\;\left(\Sigma_{ab},J_a\right)\;,\nonumber\\
P_A&\equiv&\left(P_a,P_0\right)\;\;=\;\;\left(P_a,T\right)\;,\label{Poinc2a}
\end{eqnarray}
with lowercase Latin indices running as $a,b,c,\dots,h\in\{1,2,3\}$. The algebra \eqref{Poinc1} decomposes as
\begin{eqnarray}
\left[\Sigma_{ab},\Sigma_{cd}\right]&=&\frac{1}{2}\left(\delta_{ad}\Sigma_{bc}-\delta_{ac}\Sigma_{bd}+\delta_{bc}\Sigma_{ad}-\delta_{bd}\Sigma_{ac}\right)\;,\nonumber\\
\left[\Sigma_{ab},J_c\right]&=&\frac{1}{2}\left(\delta_{bc}J_a-\delta_{ac}J_b\right)\;,\nonumber\\
\left[J_a,J_b\right]&=&-\frac{1}{2}\Sigma_{ab}\;,\nonumber\\
\left[\Sigma_{ab},P_c\right]&=&\frac{1}{2}\left(\delta_{bc}P_a-\delta_{ac}P_b\right)\;,\nonumber\\
\left[J_a,P_b\right]&=&-\frac{1}{2}\delta_{ab}T\;,\nonumber\\
\left[J_a,T\right]&=&\frac{1}{2}P_a\;,\label{Poinc2b}
\end{eqnarray}
and zero otherwise. The consequences for the fields are
\begin{eqnarray}
E&=&e^aP_a+qT\;,\nonumber\\
Y&=&\omega^{ab}\Sigma_{ab}+\theta^aJ_a\;,\label{fields1}
\end{eqnarray}
where $e^a$, called here simply by \emph{space vierbein}, relates coordinates in spacetime $M$ with coordinates in the tangent space of 3-manifolds $T_3(M)$. The time-projected vierbein, here called \emph{time vierbein} $q$, connects coordinates in spacetime with $T(M)/T_3(M)$. In the same spirit, $\omega^{ab}$ is called \emph{spin connection} while $\theta^a$ is the \emph{boost connection}. For the field strengths we have
\begin{eqnarray}
W&=&T^aP_a+\mathcal{Q}T\;,\nonumber\\
\Omega&=&\Omega^{ab}\Sigma_{ab}+S^aJ_a\;,\label{fields2}
\end{eqnarray}
with
\begin{eqnarray}
T^a&=&De^a-\frac{1}{2}q\theta^a\;\;=\;\;K^a-\frac{1}{2}q\theta^a\;,\nonumber\\
\mathcal{Q}&=&dq-\frac{1}{2}\theta_ae^a\;=\;Q-\frac{1}{2}\theta_ae^a\;,\label{T2}
\end{eqnarray}
and
\begin{eqnarray}
\Omega^a_{\phantom{a}b}&=&d\omega^a_{\phantom{a}b}+\omega^a_{\phantom{a}c}\omega^c_{\phantom{c}b}-\frac{1}{4}\theta^a\theta_b\;\;=\;\;R^a_{\phantom{a}b}-\frac{1}{4}\theta^a\theta_b\;,\nonumber\\
S^a&=&D\theta^a\;.\label{curv2}
\end{eqnarray}
In the above expressions, the covariant derivative $D\cdot\equiv d\cdot+[\omega,\cdot]$ is taken with respect to the $SO(3)$ sector. The field $R^{ab}$ will be called \emph{space curvature} while $S^a$ is the \emph{boost curvature}. The fields $T^a$, $K^a$, and Q are named, respectively, \emph{space torsion}, \emph{reduced torsion}, and \emph{time torsion}.

The non-relativistic limit of the Poincar\'e group is obtained following reference \cite{Bergshoeff:2017btm}. First, the boost and temporal translational generators are redefined by
\begin{eqnarray}
J_a&=&\zeta G_a\;,\nonumber\\
T&=&\zeta^{-1}H\;,\label{resc1}
\end{eqnarray}
together with the rescalings
\begin{eqnarray}
\theta^a&\longmapsto&\zeta^{-1}\theta^a\;,\nonumber\\
q&\longmapsto&\zeta q\;,\label{resc2}
\end{eqnarray}
in such a way that definitions \eqref{fields1} remain unchanged. At this point, the limit $\zeta\longrightarrow\infty$, which is equivalent to consider $1/c\longrightarrow0$, can be performed. The consequence for the Poincar\'e algebra \eqref{Poinc2b}, at leading order, is that it is contracted down to the Galilei algebra, namely,  
\begin{eqnarray}
\left[\Sigma_{ab},\Sigma_{cd}\right]&=&\frac{1}{2}\left(\delta_{ad}\Sigma_{bc}-\delta_{ac}\Sigma_{bd}+\delta_{bc}\Sigma_{ad}-\delta_{bd}\Sigma_{ac}\right)\;,\nonumber\\
\left[\Sigma_{ab},P_c\right]&=&\frac{1}{2}\left(\delta_{bc}P_a-\delta_{ac}P_b\right)\;,\nonumber\\
\left[\Sigma_{ab},G_c\right]&=&\frac{1}{2}\left(\delta_{bc}G_a-\delta_{ac}G_b\right)\;,\nonumber\\
\left[G_a,H\right]&=&\frac{1}{2}P_a\;,\label{gal1}
\end{eqnarray}
and zero otherwise. Thence, the limit implies on an In\"on\"u-Wigner contraction \cite{Inonu:1953sp} of the form $ISO(1,3)\longrightarrow G(1,3)=SO(3)\times B(3)\times \mathbb{R}_s^3\times \mathbb{R}_t$ with $B(3)$ being the Galilean boosts, see for instance \cite{Bergshoeff:2017btm,Bergshoeff:2019ctr}. From the algebra \eqref{gal1}, one easily observes that the Galilei group is not a semi-simple Lie group since the annexes $B(3)\times \mathbb{R}_s^3$ and $\mathbb{R}_t$ are normal Abelian subgroups of the Galilei group. Hence, there is no invariant Killing form. In fact, the isometries of the Galilei group imply on degenerate metric tensors:
    \begin{eqnarray}
    \eta_t&\equiv&\mathrm{diag}(1,0,0,0)\;,\nonumber\\
    \eta_s&\equiv&\mathrm{diag}(0,1,1,1)\;.\label{metric1}
    \end{eqnarray}
Clearly, $\eta_s\equiv\delta_{ab}$ is the 3-dimensional Euclidean flat metric while $\eta_t$ is its equivalent at time axis. These metrics are orthogonal to each other. The local invariant metrics \eqref{metric1} induces two sets of metrics on $\mathbb{\mathrm{M}}$,
\begin{eqnarray}
\tau_{\mu\nu}&=&q_\mu q_\nu\;,\nonumber\\
g_{\mu\nu}&=&e^a_\mu e^b_\nu \delta_{ab}\;.\label{metric2}
\end{eqnarray}
and
\begin{eqnarray}
\tau^{\mu\nu}&=&q^\mu q^\nu\;,\nonumber\\
g^{\mu\nu}&=&e_a^\mu e_b^\nu\;
\delta^{ab}\;.\label{metric2a}
\end{eqnarray}
Moreover, due to the relations \eqref{inv0} and \eqref{metrics0}, the following inverse relations hold
\begin{eqnarray}
e^a_\mu e_b^\mu&=&\delta^a_b\;,\nonumber\\
q^\mu q_\mu&=&1\;,\nonumber\\
q^\mu e^a_\mu&=&q_\mu e_a^\mu\;=\;0\;,\nonumber\\
e^a_\mu e_a^\nu&=&\delta_\mu^\nu-q_\mu q^\nu\;.\label{inv1}
\end{eqnarray}
Metrics \eqref{metric2} and \eqref{metric2a} relate through
\begin{eqnarray}
    \tau_{\mu\alpha}\tau^{\alpha\nu}&=&q_\mu q^\nu\;,\nonumber\\
    g_{\mu\alpha}g^{\alpha\nu}&=&\delta_\mu^\nu-q_\mu q^\nu\;.
\end{eqnarray}

The fields \eqref{fields1} remains unchanged but the fields \eqref{T2} and \eqref{curv2} reduce to
\begin{eqnarray}
T^a&=&K^a-\frac{1}{2}q\theta^a\;,\nonumber\\
Q&=&dq\;,\nonumber\\
\Omega^a_{\phantom{a}b}&=&R^a_{\phantom{a}b}\;,\nonumber\\
S^a&=&D\theta^a\;.\label{curv3}
\end{eqnarray}

The infinitesimal gauge transformations \eqref{gt0} of the fields are contracted down to\footnote{The rescaling $\alpha^{a0}\longrightarrow\zeta^{-1}\alpha^{a0}$ must be taken.} Galilei gauge transformations, 
\begin{eqnarray}
\delta\omega^a_{\phantom{a}b}&=&D\alpha^a_{\phantom{a}b}\;,\nonumber\\
\delta\theta^a&=&D\alpha^a-\alpha^a_{\phantom{a}b}\theta^b\;,\nonumber\\
\delta e^a&=&-\alpha^a_{\phantom{a}b}e^b-\frac{1}{2}\alpha^aq\;,\nonumber\\
\delta q&=&0\;,\label{gt1}
\end{eqnarray}
where $\alpha=\alpha_{ab}\Sigma_{ab}+\alpha^aG_a$. For completeness, we write down the gauge transformations of curvatures and torsions,
\begin{eqnarray}
\delta R^a_{\phantom{a}b}&=&R^a_{\phantom{a}c}\alpha^c_{\phantom{c}b}-R_b^{\phantom{b}c}\alpha_c^{\phantom{c}a}\;,\nonumber\\
\delta S^a&=&S^b\alpha_b^{\phantom{b}a}+R^a_{\phantom{a}b}\alpha^b\;,\nonumber\\
\delta T^a&=&T^b\alpha_b^{\phantom{b}a}-\frac{1}{2}Q\alpha^a\;,\nonumber\\
\delta Q&=&0\;.\label{gt2}
\end{eqnarray}
Up to tautological relations, the hierarchy relations \eqref{hier1} lead to the following set of self-consistent equations
\begin{eqnarray}
dQ&=&0\;,\nonumber\\
DK^a&=&R^a_{\phantom{a}b}e^b\;,\nonumber\\
DS^a&=&R^a_{\phantom{a}b}\theta^b\;,\nonumber\\
DR^a_{\phantom{a}b}&=&0\;.\label{Bia2}
\end{eqnarray}

The consequences of the In\"on\"u-Wigner contraction setup here discussed are the basis to study the non-relativistic limit of gravity theories. We begin with the EH action as a first example in the following section.

\section{Galilei gravity}\label{EH}

What we call in this work \emph{Galilei gravity} can be obtained from the non-relativistic limit of the Einstein-Hilbert action \eqref{EH} at leading order in the $1/c$ expansion. This task is achieved by considering the In\"on\"u-Wigner contraction of the Poincar\'e group to the Galilei group, as described in Sect.~\ref{IW}. See for instance \cite{Bergshoeff:2017btm}. Thence, employing decompositions \eqref{fields1} and \eqref{fields2}, the Einstein-Hilbert action \eqref{ehaction0} reads
\begin{equation}
S_{EH}=\kappa\int\epsilon_{abc}\left(2qR^{ab}e^c-\frac{1}{2}q\theta^a\theta^be^c-S^ae^be^c\right)+S_m\;,\label{ehaction1}
\end{equation}
with $\epsilon_{abc}=\epsilon_{\underline{0}abc}$ where $\underline{0}$ stands for the time direction in tangent space. Performing the rescalings \eqref{resc2} we get
\begin{equation}
  S_{EH}=\kappa\int\epsilon_{abc}\left(2\zeta qR^{ab}e^c-\frac{1}{2}\zeta^{-1}q\theta^a\theta^be^c-\zeta^{-1}S^ae^be^c\right)+S_m\;.\label{ehaction2}
\end{equation}
The non-relativistic limit is then obtained by taking $\zeta\longrightarrow\infty$ together with the coupling rescaling $\kappa\longmapsto\zeta^{-1}\kappa$. The result is the Galilei gravity action \cite{Bergshoeff:2017btm,Bergshoeff:2019ctr},
\begin{equation}
  S_{G}=2\kappa\int\epsilon_{abc}qR^{ab}e^c+S_m\;.\label{galaction0}
\end{equation}
In this procedure, we assume that the corresponding limit of the matter action $S_m$ is consistent. 

It is worth mentioning that the pure gravitational sector of action \eqref{galaction0}, $S_{pG}=S_G-S_m$, is invariant under local scale invariance \cite{Bergshoeff:2017btm,Bergshoeff:2019ctr}
\begin{eqnarray}
q&\longmapsto&\exp{\left(-\phi\right)}q\;,\nonumber\\
e^a&\longmapsto&\exp{\left(\phi \right)}e^a\;,\label{conf1}
\end{eqnarray}
with $\phi=\phi(x)$ being a local gauge parameter. In functional form, symmetry \eqref{conf1} implies on
\begin{eqnarray}
e^a\frac{\delta S_{pG}}{\delta e^a}-q\frac{\delta S_{pG}}{\delta q}=0\;.\label{conf2}
\end{eqnarray}
Identity \eqref{conf2} assigns a global charge $\pm1$ for $e$ and $q$, respectively. 

The field equations can be easily computed from the action \eqref{galaction0} for the fields $q$, $e$, $\omega$, and $\theta$, respectively given by
\begin{eqnarray}
\epsilon_{abc}R^{ab}e^c&=&-\frac{1}{2\kappa}\frac{\delta S_m}{\delta q}\;,\nonumber\\
\epsilon_{abc}qR^{bc}&=&\frac{1}{2\kappa}\frac{\delta S_m}{\delta e^a}\;,\nonumber\\
\epsilon_{abc}\left(Qe^c-qK^c\right)&=&-\frac{1}{2\kappa}\frac{\delta S_m}{\delta\omega^{ab}}\;,\nonumber\\
0&=&\frac{\delta S_m}{\delta \theta^a}\;.\label{feqG1}
\end{eqnarray}
The last equation implies on two important features of Galilei gravity: First, it establishes a constraint saying that the non-relativistic limit of the matter content must not couple with the boost connection; And second, that the field $\theta^a$ remains free, since it does not appear in the field equations (nor in the action \eqref{galaction0}). Typically, this problem is solved by considering higher order corrections in $\zeta^{-1}$ at the non-relativistic limit of the Einstein-Hilbert action \eqref{ehaction1}. Otherwise, a constraint for $\theta^a$ must be implemented by hand. The second equation in \eqref{feqG1} implies on another constraint on the matter content, namely,
\begin{equation}
    q\frac{\delta S_m}{\delta e^a}=0\;.\label{mattconst1}
\end{equation}
Moreover, combining the first and second equations in \eqref{feqG1} we get
\begin{eqnarray}
e^a\frac{\delta S_m}{\delta e^a}-q\frac{\delta S_m}{\delta q}=0\;,\label{mattconst2}
\end{eqnarray}
saying that symmetry \eqref{conf1} must be obeyed by the non-relativistic limit of the matter content as well. Constraint \eqref{mattconst1} states that whenever $e^a$ appears at the matter action, a $q$ must be there as well, \emph{i.e.}, $e^a$ and $q$ always come in pairs $qe^a$. Constraint \eqref{mattconst2} reinforces constraint \eqref{mattconst1} by saying that the matter action must be linear in $e^a$.

\subsection{Vacuum solutions}

The field equations \eqref{feqG1} in vacuum are of particular interest. The first two equations imply on
\begin{equation}
    R^{ab}_{\mu\nu}e^\mu_ce^\nu_d=R^{ab}_{cd}=0\;.\label{R0}    
\end{equation}
The third equation, assumes the form
\begin{equation}
    Qe^a-qK^a=0\;.\label{Q1}
\end{equation}
We focus in this equation to complement the curvature solution \eqref{R0} with possible torsional solutions in vacuum (other than the trivial inconsistent solution $q=0$). It is important to keep in mind that boost connection and boost curvatures remain free.

\subsubsection{A no-go constrained solution}

As a no-go result, let $e^a$ and $q$ be related by
\begin{equation}
e^a=qn^a\;,\label{eqn}
\end{equation}
with $n^a$ being an arbitrary algebra-valued 0-form. This tentative ansatz is a solution of equation \eqref{Q1}, as one can easily check. Nevertheless, such attempt is inconsistent with the degenerate nature of the metric. For instance, this is evident by contracting the ansatz \eqref{eqn} with $q^\mu$, resulting in $q_\mu q^\mu=0$, which contradicts relations \eqref{inv1}. In other words, $q$ and $e^a$ are orthogonal by definition while the proposal \eqref{eqn} enforces these fields to be parallel.

\subsubsection{Twistless torsion solution}

In tensor notation, equation \eqref{Q1} can be manipulated to provide the following geometric relations
\begin{eqnarray}
e^\mu_be^\nu_cQ_{\mu\nu}\;\;=\;\;Q_{bc}&=&0\;,\nonumber\\
2q^\mu Q_{\mu\nu} e^\nu_b+K^a_{\alpha\nu}e^\nu_be^\alpha_a\;\;=\;\;
     2Q_{\underline{0}b}+K^a_{ab}&=&0\;,\nonumber\\
  \frac{1}{2} \left(K^d_{db}\delta^a_c-K^d_{dc}\delta^a_b\right)+K^a_{bc}&=&0\;.\label{Q2}
\end{eqnarray}
The first equation in \eqref{Q2} is the well known twistless torsion condition. It fixes a specific spacetime foliation for which time torsion $Q$ has no projection on the locally inertial leaves, ensuring thus causality. This condition also allows the implementation of Newtonian absolute (which means that the first equation in \eqref{Q2} does not imply on absolute Newtonian time. But, it allows its implementation by choice. See also Section \ref{STTS}). Equations \eqref{R0} and \eqref{Q2} synthesize the well-known results of Galilei gravity in vacuum \cite{Christensen:2013rfa,Bergshoeff:2017btm}. This solution is known as \emph{twistless torsion solution} and it still gives room for non-vanishing time torsion solutions. Notice that, imposition of $K^a_{bc}=0$ leads to $Q_{\underline{0}a}=0$ in tangent projections. Moreover, $Q_{ab}=Q_{\underline{0}a}=0$ does not imply on $Q=0$. The latter constitutes the \emph{torsionfree condition}, which we discuss in the next section.

\subsubsection{Torsionfree solution}\label{STTS}

As just mentioned, \emph{torsionfree solution} is given by
    \begin{equation}
        Q=0\;,\label{Q0}
    \end{equation}
which is clearly a stronger condition than the first equation in \eqref{Q2}. The consequence of the torsionfree condition \eqref{Q0} is that the time vierbein can be chosen as an exact field, $q=dT$. Thus, time vierbein is defined by an arbitrary scalar function $T(x)$ which, among an infinity number of possibilities, can be fixed as the time coordinate $t$. Since $q$ is gauge invariant, this solution is absolute up to time translations $t\longmapsto t+\mathrm{constant}$. Hence, $t$ is identified with the absolute Newtonian time. If no other assumption is made over $q$, equation \eqref{Q1} is satisfied for 
\begin{equation}
    K^a=0\;.\label{k0}
\end{equation}
If $S^a$ remains free, equations \eqref{Q0} and \eqref{k0} compose a torsional solution case, $T^a\propto q\theta^a$. For a torsionless solution, one might choose $\theta^a=0\;\longrightarrow\;S^a=0$. Nevertheless, \eqref{k0} is not a requirement. Equation \eqref{Q1} together with \eqref{Q0} imply on the softer condition for the reduced torsion
\begin{equation}
    e^\mu_be^\nu_cK^a_{\mu\nu}=K^a_{bc}=0\;,\label{k1}
\end{equation}
with $\theta^a$ and $S^a$ still free to be fixed.

\subsubsection{A Weitzenb\"ock-like solution}
    
An \emph{ansatz} for $Q$ can be made,
\begin{equation}
   Q=qn\;\;\Rightarrow\;\;(d+n)q=0\;,\label{Qq1}
\end{equation} 
with $n$ being an arbitrary gauge invariant 1-form field. Combining \eqref{Q1} with \eqref{Qq1} provides $q(ne^a-K^a)=0$, which is satisfied, for instance, by
\begin{equation}
    K^a=ne^a\;\;\Rightarrow\;\;(D-n)e^a=0\;.\label{Kne1}
\end{equation}
The first Bianchi identy in \eqref{Bia2} states that $qn$ must be closed,
\begin{equation}
    d(qn)=0\;.\label{closed1}
\end{equation}
Using now the second Bianchi identity of \eqref{Bia2}, we find the curvature associated to the proposed solution \eqref{Qq1} and \eqref{Kne1},
\begin{equation}
    R^a_{\phantom{a}b}=dn\delta^a_b\;,\label{R0a}
\end{equation}
which is inconsistent due to the symmetry properties of the indices $a$ and $b$. Hence, inevitably, space curvature must vanish and $n$ must be closed
\begin{equation}
    R^a_{\phantom{a}b}=0\;\;\Rightarrow\;\;dn=0\;.\label{R0b}
\end{equation}
Combination of equations \eqref{closed1} and \eqref{R0b} implies on the torsionfree solution \eqref{Q0}. Hence, $q$ must be parallel to $n$. Condition $Q=0$ together with equations \eqref{Kne1} and \eqref{R0b} define a torsionfree solution with vanishing space curvature and non-vanishing reduced torsion. Hence, spacetime will be a Weitzenb\"ock-like space\footnote{In GR, Weitzenb\"ock spacetimes serve as scenario for the so called Teleparallel Gravity theories. See, for instance \cite{Garecki:2010jj,Aldrovandi:2013wha,Golovnev:2018red} and references therein.} \cite{Weitzenbock:1923boo} if we also demand $S^a=0$. 

In the particular case where $n$ is a constant field, equation \eqref{Qq1} is easily solved for
\begin{equation}
    q=a\exp(-n_\mu x^\mu)+bn\;,\label{nconst1}
\end{equation}
where $a$ is another constant 1-form field and $b$ a constant scalar field. It can be easily checked that, with the help of equation \eqref{R0b}, expression \eqref{nconst1} satisfies equation \eqref{closed1}. To be even more specific in this example, we consider ADM formalism in the temporal gauge \cite{Misner:1974qy}. In the first order formalism this means that $E^0_i=0$ and $E^0_0=N$, with $N$ being the lapse function. Then, $q=E^0_0dt+E^0_idx^i=Ndt$, with $t$ being chosen as the Newtonian absolute time whose existence is ensured by the torsionfree condition $Q=0$. Also, choosing\footnote{In that case, $n$ itself can be used to define a consistent foliation with $n$ being dual to the normal vector of the spatial leaves.} $n=\mathsf{n}dt$ and $a=\mathsf{a}dt$, with $\mathsf{n}$ and $\mathsf{a}$ being positive constants, we get
\begin{equation}
    N=\mathsf{a}\exp(-\mathsf{n}t)+b\mathsf{n}\;.\label{nconst2}
\end{equation}
Since $N=N(t)\ne1$, the lapse function characterizes a non-trivial time different from the absolute time $t$. On the other hand, \eqref{nconst2} also says that $q\longrightarrow bn$ as time evolves. Hence, $b\mathsf{n}$ is the asymptotic lapse function. We can set then $b=\mathsf{n}^{-1}$ in order to identify the asymptotic time with the Newtonian absolute time where $N_{abs}=1$. Hence, 
\begin{equation}
    N=\mathsf{a}\exp(-\mathsf{n}t)+1\;.\label{nconst3}
\end{equation}
Notice that for this specific case, $Q=0$ because $q$ is parallel to $n$, even though $q\ne dt$, except asymptotically. Nevertheless, it is possible to write $q=dT$ where
\begin{equation}
    T=-\frac{\mathsf{a}}{\mathsf{n}}\exp(-\mathsf{n}t)+t+c\;,\label{T0}
\end{equation}
with $c$ a constant to be fixed. In fact, if we interpret $T$ as a non-Newtonian time, we can fix $c$ by demanding $T$ to be positive definite in the domain $t\in[0,\infty]$. Since $T\in[-\mathsf{a}/\mathsf{n}+c,\infty)$, for $T\ge0$ we get that $c\ge\mathsf{a}/\mathsf{n}$. Thence, we can set
\begin{equation}
     T=\frac{\mathsf{a}}{\mathsf{n}}\left[1-\exp(-\mathsf{n}t)\right]+t\;,\label{T1}
\end{equation}
so we have the range $T\in[0,\infty)$. Note that $T(t)\Big|_{t\rightarrow\infty}\sim t+\mathsf{a}/\mathsf{n}$. Moreover, $N=dT/dt$ characterizes the time flow rate with respect to the Newtonian time. Thus, non-Newtonian and Newtonian time intervals, respectively $\Delta T=T_2-T_1$ and $\Delta t=t_2-t_1$, are related by
\begin{equation}
    \Delta T=\int_{t_1}^{t_2}N(t)dt=T\Big|_{t_1}^{t_2}=-\frac{\mathsf{a}}{\mathsf{n}}\exp(-\mathsf{n}t_2)+\frac{\mathsf{a}}{\mathsf{n}}\exp(-\mathsf{n}t_1)+\Delta t\approx \left(1+\mathsf{a}\right)\Delta t+0(\mathsf{n})\;.
\end{equation}
This is an explicit example of time dilation in NC gravity. A nice discussion about time dilation in NC gravity can be found in \cite{Hansen:2018ofj}.

Finally, still in this example, space torsion reads (see \eqref{Kne1})
\begin{equation}
    K^a=\mathsf{n}e_i^adtdx^i\;,\label{K0}
\end{equation}
with lowercase Latin indices $i,j,\ldots,s$ running through $\{1,2,3\}$. Thence, while $R^{ab}_{\mu\nu}=S^a_{\mu\nu}=Q_{\mu\nu}=K^a_{ij}=0$, we have $K^a_{0i}\ne0$. The non-triviality of spacetime geometry is encoded in \eqref{K0}. We remark that we can easily set the Newtonian absolute time by choosing $\mathsf{a}=0$. Nevertheless, expression \eqref{K0} remains valid since it does not depend on the 1-form $a$. To find the components of $e^a_i$, we can set $\omega^{ab}=0$, which solves \eqref{R0} \emph{\`a la} Weitzenb\"ock while giving room for torsional degrees of freedom. Thence, equation \eqref{K0} can be directly integrated to provide
\begin{equation}
    e^a=c^a_i\exp(\mathsf{n}t)dx^i\;,\label{vier0}
\end{equation}
with $c^a=c^a_idx^i$ being a constant 1-form field. Relations \eqref{inv1} allow to construct the full vierbein solution whose non-vanishing components are
\begin{eqnarray}
q_0&=&N\;,\nonumber\\
q^0&=&N^{-1}\;,\nonumber\\
e^a_i&=&\frac{\mathsf{a}}{(N-1)}c^a_i\;,\nonumber\\
e_a^i&=&\frac{(N-1)}{\mathsf{a}}c_a^i\;,\label{vier1}
\end{eqnarray}
with
\begin{eqnarray}
    c_a^ic^b_i&=&\delta_a^b\;,\nonumber\\
    c_a^ic^a_j&=&\delta_j^i\;.\label{c0}
\end{eqnarray}
It can be checked using \eqref{metric2} and \eqref{metric2a} that this solution induces conformally flat metrics of the form
\begin{eqnarray}
\tau^{00}&=&N^{-2}\;,\nonumber\\
\tau^{ij}&=&0\;,\nonumber\\
g^{ij}&=&\frac{(N-1)^2}{\mathsf{a}^2}\delta^{ij}\;,\nonumber\\
g^{00}&=&0\;,\label{metric3a}
\end{eqnarray}
and
\begin{eqnarray}
\tau_{00}&=&N^2\;,\nonumber\\
\tau_{ij}&=&0\;,\nonumber\\
g_{ij}&=&\frac{\mathsf{a}^2}{(N-1)^2}\delta_{ij}\;,\nonumber\\
g_{00}&=&0\;.\label{metric3b}
\end{eqnarray}
One can reach an Euclidean three-dimensional metric by performing a conformal transformation while respecting the symmetry \eqref{conf1}. To see this property, we perform the following transformations, which keep the action \eqref{galaction0} invariant,
\begin{eqnarray}
e^a_i&\longrightarrow&\exp(-\mathsf{n}t)e^a_i\;,\nonumber\\
q&\longrightarrow&\exp(\mathsf{n}t)q\;.
\end{eqnarray}
Henceforth, the space metric trivializes while the new temporal metric reads
\begin{eqnarray}
\tau^{00}&=&\frac{(1-N^{-1})^2}{\mathsf{a}^2}\;,\nonumber\\
\tau_{00}&=&\frac{\mathsf{a}^2}{(1-N^{-1})^2}\;.
\end{eqnarray}
One can also trivialize temporal metric while keeping space metric conformally flat by performing an appropriate conformal transformation. To trivialize all metrics simultaneously, Newtonian time must be evoked by setting $\mathsf{a}=0$.

\subsection{Solutions in the presence of matter}

Going back to equations \eqref{feqG1} in the presence of matter, we rewrite them as
\begin{eqnarray}
\epsilon_{abc}R^{ab}e^c&=&-\frac{1}{2\kappa}\tau\;,\nonumber\\
qR^{ab}&=&\frac{1}{4\kappa}\epsilon^{abc}\tau_c\;,\nonumber\\
Qe^a-qK^a&=&-\frac{1}{4\kappa}\epsilon^{abc}\sigma_{bc}\;,\nonumber\\
0&=&\sigma_a\;,\label{feqG2}
\end{eqnarray}
with the following definitions
\begin{equation}
\tau=\frac{\delta S_m}{\delta q}\;,\;\;
\tau_a=\frac{\delta S_m}{\delta e^a}\;,\;\;
\sigma_{ab}=\frac{\delta S_m}{\delta\omega^{ab}}\;,\;\;
\sigma_a=\frac{\delta S_m}{\delta \theta^a}\;,\label{em1}
\end{equation}
with $\tau$ and $\tau_a$ being 3-forms associated with the energy-momentum tensor of the matter content while $\sigma_a$ and $\sigma_{ab}$ are 3-forms describing the spin-density of the matter content. In terms of $\tau$ and $\tau_a$, constraints \eqref{mattconst1} and \eqref{mattconst2} are rewritten as
\begin{eqnarray}
q\tau_a&=&0\;,\nonumber\\
e^a\tau_a-q\tau&=&0\;.\label{em3}
\end{eqnarray}
These equations suggest that $\tau$ and $\tau_a$ can be chosen as
\begin{eqnarray}
\tau&=&-4e^am_a\;,\nonumber\\
\tau_a&=&4qm_a\;,\label{em4}
\end{eqnarray}
with $m_a$ being a 2-form describing the matter content. The factor $4$ is just a convenient normalization factor. We notice that $m_a$ does not depend on $q$ nor $e^a$ (See the argument below \eqref{mattconst2}). Moreover, there is no need in defining $m$, the $m_a$ time counterpart, since the second of \eqref{em3} (which is just \eqref{mattconst2}) defines a constraint between $\tau$ and $\tau_a$. The matter content is then written as\footnote{We remark that this form of the matter action is not general. It works fine for external sources. In the case of dynamical fields, the Hodge star usually enters in the game. Moreover, even if the matter action $S_m$ does not respect the symmetry \eqref{conf1} explicitly, the symmetry will then enforce a constraint between the matter fields.}
\begin{equation}
S_m=-4\int qe^am_a\;.\label{matter0}
\end{equation}
Thus,
\begin{equation}
    \sigma_{ab}=-4qe^c\frac{\delta m_c}{\delta\omega^{ab}}\;,\;\;
\frac{\delta m_b}{\delta \theta^a}=0\;.\label{sigma0}
\end{equation}
Imposing torsionfree condition \eqref{Q0}, applying $D$ on the second of \eqref{feqG2}, and using Bianchi identities \eqref{Bia2}, one also finds
\begin{equation}
    D\tau_a=0\;.\label{em2}
\end{equation}
The field equations \eqref{feqG2} read now
\begin{eqnarray}
\epsilon_{abc}R^{ab}e^c&=&\frac{2}{\kappa}e^am_a\;,\nonumber\\
qR^{ab}&=&\frac{1}{\kappa}q\epsilon^{abc}m_c\;,\nonumber\\
qK^a&=&-\frac{1}{\kappa}\epsilon^{abc}qe^d\frac{\delta m_d}{\delta\omega^{bc}}\;.\label{feqG3}
\end{eqnarray}
Clearly, reduced torsionless geometry requires that the matter content does not depend on the spin connection or, at least, that $K^a$ has a part independent of $q$. For fully torsionless geometry, one also has to set $\theta^a=0$, see \eqref{T2}. The first two equations in expressions \eqref{feqG3} are satisfied for
\begin{equation}
    R^{ab}=\frac{1}{\kappa}\epsilon^{abc}m_c\;.\label{sol1}
\end{equation}
The third equation admits the solution
\begin{equation}
K^a=-\frac{1}{\kappa}\epsilon^{abc}e^ds_{bcd}\;,\label{sol2}
\end{equation}
with $s_{abc}=\delta m_c/\delta\omega^{ab}$ being a 1-form related to the spin-density matter distribution. It is more convenient to define the 2-form $s_{ab}=-e^cs_{abc}$. Thence,
\begin{equation}
K^a=\frac{1}{\kappa}\epsilon^{abc}s_{bc}\;.\label{sol2a}
\end{equation}
Equation \eqref{sol2a} and the Bianchi identity \eqref{Bia2} imply on the extra constraint for the matter action $S_m$
\begin{equation}
Ds_{ab}=\frac{1}{2}\left(e_am_b-e_bm_a\right)\;.    \label{Ds1}
\end{equation}

It remains to fix $S^a$. Field equations say nothing about it. But the Bianchi identities \eqref{Bia2} imply on
\begin{equation}
DS^a=\frac{1}{\kappa}\epsilon^{abc}m_c\theta_b\;.\label{Bia4}
\end{equation}
We can fix $\theta^a$ or $S^a$ at will as long as \eqref{Bia4} is respected. For instance, the trivial solution $\theta^a=0\Rightarrow S^a=0$ is allowed. Non-trivially, we may set, for example,
\begin{equation}
    S^a=a\epsilon^{abc}s_{bc}\;,\label{sol3}
\end{equation}
where $a\in\mathbb{R}$. Thus, from \eqref{sol2a}, we get the constraint
\begin{equation}
    \theta^a=\kappa a\left(e^a+\bar{\theta}^a\right)\;\;\Rightarrow\;\;S^a=\kappa aK^a\;.\label{sol3a}
\end{equation}
The field $\bar{\theta}^a$ is introduced because $\theta^a$ and $e^a$ transform differently under Galilei gauge transformations. Thus, $\bar{\theta}^a$ must obey the following properties,
\begin{eqnarray}
D\bar{\theta}^a&=&0\;,\nonumber\\
\delta\bar{\theta}^a&=&-\delta e^a+\frac{1}{\kappa a}\delta\theta^a\;\;=\;\;-\alpha^a_{\phantom{a}b}\bar{\theta}^b+\frac{1}{2}\alpha^aq+\frac{1}{\kappa a}D\alpha^a\;.\label{prop1}
\end{eqnarray} 
Thence, up to an arbitrary dimensionless non-vanishing real constant $a$, $\theta^a$ is totally fixed. Obviously, $a=0$ sets $\theta^a=0$ and $S^a=0$.

The main conclusion of this section is that Galilei gravity accepts consistent solutions in the presence of matter within torsionfree solution ($Q=0$ - which allows, as mentioned before, to employ absolute Newtonian time by setting $q=dt$) while having non-trivial curvatures and reduced torsion.

\section{Galilei-Cartan gravity}\label{GC}

We now consider the non-relativistic limit of the Mardones-Zanelli action \eqref{mzaction1}. Decompositions \eqref{fields1} and \eqref{fields2}, imply on
\begin{eqnarray}
  S_{MZ}&=&\kappa\int\left[\epsilon_{abc}\left(2qR^{ab}e^c-\frac{1}{2}q\theta^a\theta^be^c-S^ae^be^c\right)+2\Lambda\epsilon_{abc}qe^ae^be^c\right]+\nonumber\\
  &+&\int\left[z_1\left(R^{ab}e_ae_b-qe_aS^a+\frac{1}{4}e_a\theta^ae_b\theta^b\right)+z_2\left(Q^2+K^aK_a-q\theta^aK_a+e_a\theta^aQ+\frac{1}{4}e_a\theta^ae_b\theta^b\right)\right]+\nonumber\\
    &-&\int\left[2z_3\epsilon_{abc}S^a\left(R^{bc}-\frac{1}{4}\theta^b\theta^c\right)-z_4\left(R^{ab}R_{ab}+\frac{1}{2}S^aS_a-\frac{1}{2}R^{ab}\theta_a\theta_b\right)\right]+S_m\;.\label{mzaction2}
\end{eqnarray}
From the rescaling of the fields \eqref{resc2}, we get
\begin{eqnarray}
  S_{MZ}&=&\kappa\int\left[\epsilon_{abc}\left(2\zeta qR^{ab}e^c-\frac{1}{2}\zeta^{-1}q\theta^a\theta^be^c-\zeta^{-1}S^ae^be^c\right)+2\Lambda\zeta\epsilon_{abc}qe^ae^be^c\right]+\;,\nonumber\\
  &+&\int\left[z_1\left(R^{ab}e_ae_b-qe_aS^a+\frac{1}{4}\zeta^{-2}e_a\theta^ae_b\theta^b\right)+\right.\nonumber\\
  &+&\left.z_2\left(\zeta^2Q^2+K^aK_a-q\theta^aK_a+e_a\theta^aQ+\frac{1}{4}\zeta^{-2}e_a\theta^ae_b\theta^b\right)\right]+\nonumber\\
    &-&\int\left[2z_3\epsilon_{abc}\left(\zeta^{-1}S^aR^{bc}-\frac{1}{4}\zeta^{-3}S^a\theta^b\theta^c\right)-z_4\left(R^{ab}R_{ab}+\frac{1}{2}\zeta^{-2}S^aS_a-\frac{1}{2}\zeta^{-2}R^{ab}\theta_a\theta_b\right)\right]+\nonumber\\
    &+&S_m\;.\label{galaction1}
\end{eqnarray}
The corresponding non-relativistic limit is obtained by taking $\zeta\longrightarrow\infty$ together with the coupling rescalings\footnote{In \eqref{resc3}, the parameters were modified, if needed, in order to eliminate potential divergences at the limit $\zeta\longrightarrow\infty$. Except for the cosmological constant, where we followed the scaling employed in \cite{Aldrovandi:1998im,Hansen:2020pqs}. One could wonder a possible role of the cosmological constant in the non-relativistic limit of the Mardones-Zanelli action and try a different scale for it. Nevertheless, if this scaling would not be employed, the resulting non-relativistic theory would be inconsistent at leading order--See Appendix \eqref{CCap}.}
\begin{eqnarray}
\kappa&\longmapsto&\zeta^{-1}\kappa\;,\nonumber\\
\Lambda&\longmapsto&\zeta^{-2}\Lambda\;,\nonumber\\
z_1&\longmapsto&z_1\;,\nonumber\\
z_2&\longmapsto&\zeta^{-2}z_2\;,\nonumber\\
z_3&\longmapsto&z_3\;,\nonumber\\
z_4&\longmapsto&z_4\;.\label{resc3}
\end{eqnarray}
Thus, the MZ action \eqref{galaction1} reduces to its non-relativistic limit at leading order,
\begin{equation}
  S_{GMZ}=2\kappa\int\left[\epsilon_{abc}qR^{ab}e^c+z_1\left(R^{ab}e_ae_b-qe_aS^a\right)+z_2Q^2+z_4R^{ab}R_{ab}\right]+S_m\;.\label{galaction3}
\end{equation}
It is a straightforward exercise to check that this action is invariant under the Galilei gauge transformations \eqref{gt1} and \eqref{gt2}. For now, we will refer to this theory as \emph{Galilei-Cartan gravity}. The first term in the action \eqref{galaction3} is the same that appears in the non-relativistic limit of the Einstein-Hilbert action \eqref{galaction0}. The terms accompanied by $z_1$ are new and will affect the dynamics. Terms with factors $z_2$ and $z_4$ are of topological nature (Abelian and non-Abelian Pontryagin actions, respectively) and do not contribute to the field equations. Moreover, it is assumed that the non-relativistic limit of $S_m$ is consistent. One can readily note that transformations \eqref{conf1} do not constitute a symmetry of action \eqref{galaction3}.

The field equations generated by the action \eqref{galaction3} are\footnote{See definitions \eqref{em1}.}
\begin{eqnarray}
2\kappa\epsilon_{abc}R^{ab}e^c-z_1e_aS^a&=&-\tau\;,\nonumber\\
2\kappa\epsilon_{abc}qR^{bc}-z_1\left(2R_{ab}e^b+qS_a\right)&=&\tau_a\;,\nonumber\\
2\kappa D(\epsilon_{abc}qe^c)+z_1D(e_ae_b)-\frac{z_1}{2}q(e_a\theta_b-e_b\theta_a)&=&-\sigma_{ab}\;,\nonumber\\
z_1\left(Qe^a-qK^a\right)&=&\sigma_a\;.\label{feqmz2a}
\end{eqnarray}
We may have lost the scale symmetry \eqref{conf1}, but now we have an extra equation (the last of equations \eqref{feqmz2a}) associated with $\theta^a$. Moreover, the boost connection appears explicitly in the field equations. This is a welcome feature of the non-relativistic limit of the MZ action: The system of equations is now fully determined already at leading order. Combining the first and second equations in \eqref{feqmz2a} we get
\begin{equation}
    2z_1R^{ab}e_ae_b=q\tau-e^a\tau_a\;.\label{comb1}
\end{equation}
In the same spirit, we can combine the third and fourth equations in \eqref{feqmz2a} to get
\begin{equation}
e_b\left(K_a-\frac{1}{2}q\theta_a\right)-e_a\left(K_b-\frac{1}{2}q\theta_b\right)=-\frac{1}{z_1}\sigma_{ab}-2\frac{\kappa}{z^2_1}\epsilon_{abc}\sigma^c\;.\label{comb1c}
\end{equation}

\subsection{Newtonian absolute time}\label{NAT}

An important feature that a Newton-Cartan gravity theory must have is to accept torsionfree solutions. This means that Newtonian absolute time can be safely employed. To show that this property is well accepted by the full set of field equations \eqref{feqmz2a}, let us impose $Q=0$ to them. This imposition has no direct effect on the first two equations of the field equations \eqref{feqmz2a}. On the fourth equation, however, leads to
\begin{equation}
    qK^a=-\frac{1}{z_1}\sigma_a\;.\label{q00}
\end{equation}
This equation establishes the conditions for the theory to accept torsionfree solutions, allowing to work with the absolute Newtonian time. These conditions can be summarized by: i) In the presence of matter coupled to $\theta^a$, reduced torsion $K^a$ must not vanish and have a piece independent of $q$; ii) In the absence of the coupling between $\theta^a$ and matter, $K^a$ must vanish or, at least, to be proportional to $q$. In essence, torsionfree solutions are described by the first three equations in \eqref{feqmz2a} and equation \eqref{q00}.

\subsection{Vacuum solution}

Another important feature of a consistent non-relativistic theory is the existence of vacuum solutions. But before we try to solve equations \eqref{feqmz2a} in vacuum, let us consider the vacuum solution of the former theory, \emph{i.e.}, the maximally symmetric solution \eqref{dS0} and its non-relativistic limit. Employing decomposition \eqref{fields2} in solution \eqref{dS0} we get
\begin{eqnarray}
    \left(R_o^{ab}-\frac{1}{4}\theta^a\theta^b\right)\Sigma_{ab}&=&-\Lambda e^ae^b\Sigma_{ab}\;,\nonumber\\
    S_o^aJ_a&=&2\Lambda qe^aJ_a\;.\label{dS0a}
\end{eqnarray}
To obtain the non-relativistic limit of these relations we first perform the rescalings \eqref{resc1} and \eqref{resc2} in \eqref{dS0a},
\begin{eqnarray}
    \left(R_o^{ab}-\frac{1}{4}\zeta^{-2}\theta^a\theta^b\right)\Sigma_{ab}&=&-\zeta^{-2}\Lambda e^ae^b\Sigma_{ab}\;,\nonumber\\
    S_o^aG_a&=&2\Lambda qe^aG_a\;.\label{dS0b}
\end{eqnarray}
Taking the limit $\xi\longrightarrow\infty$ in \eqref{dS0b}, the non-relativistic limit of the maximally symmetric solution \eqref{dS0} is achieved,
\begin{eqnarray}
    R_o^{ab}&=&0\;,\nonumber\\
    S_o^a&=&2\Lambda qe^a\;,\label{vac6e}
\end{eqnarray}
with full torsionless solution $T=Q=0\;\longrightarrow\;K=q\theta/2$.

It is easy to check that, if we impose full torsionless solution to equations \eqref{feqmz2a} in vacuum, third and fourth equations are automatically satisfied. The first equation trivially accepts vanishing space curvature. The second equation is satisfied by $S^a\propto q$. Thus, a generic vacuum solution of the field equations \eqref{feqmz2a} reads
\begin{eqnarray}
    R^{ab}&=&0\;,\nonumber\\
    S^a&=&aqe^a\;,\nonumber\\
    K^a&=&\frac{1}{2}q\theta^a\;,\nonumber\\
    Q&=&0\;,\label{vac6}
\end{eqnarray}
with $a$ being a squared mass parameter. It can be easily checked that this vacuum solution satisfies the Bianchi identities \eqref{Bia2}. Obviously, if we demand that this solution agrees with \eqref{vac6e} we need to set
\begin{equation}
    a=2\Lambda\;.\label{xi2}
\end{equation}
This procedure fixes $a$ by connecting the solutions \eqref{vac6} with the original relativistic vacuum solution \eqref{dS0} of the MZ action.

In summary, the non-relativistic limit of the MZ action is actually independent of $\Lambda$; Nevertheless, $\Lambda$ could still be probed in non-relativistic limits of gravitational systems through the solution \eqref{vac6} (complemented with \eqref{xi2}). Finally, for extra discussions about the non-relativistic limit of de Sitter and anti-de Sitter spacetimes, see also \cite{Aldrovandi:1998im,Hansen:2020pqs}.

\subsection{Solutions in the presence of matter}\label{IMmz}

To find solutions of equations \eqref{feqmz2a} in the presence of matter we first constrain $S_m$ to be of the form \eqref{matter0}. Notice that, differently from the Galilean case, $m_a$ can now depend on $e$, $\omega$, and $\theta$, but not on $q$. This form of $S_m$ is a bit restrictive, but sufficient for our initial purposes in probing the existence of consistent solutions in the presence of matter, at least at formal level. The field equations \eqref{feqmz2a} reduce to\footnote{Clearly,
\begin{eqnarray}
    \sigma_{ab}&=&-4qe^c\frac{\delta m_c}{\delta\omega^{ab}}\;,\nonumber\\
    \sigma_a&=&-4qe^c\frac{\delta m_c}{\delta\theta^a}\;.
\end{eqnarray}}
\begin{eqnarray}
2\kappa\epsilon_{abc}R^{ab}e^c-z_1e_aS^a&=&4e^am_a\;,\nonumber\\
2\kappa\epsilon_{abc}qR^{bc}-z_1\left(2R_{ab}e^b+qS_a\right)&=&4qm_a-4qe^b\frac{\delta m_b}{\delta e^a}\;,\nonumber\\
2\kappa D(\epsilon_{abc}qe^c)+z_1D(e_ae_b)-\frac{z_1}{2}q(e_a\theta_b-e_b\theta_a)&=&4qe^c\frac{\delta m_c}{\delta\omega^{ab}}\;,\nonumber\\
z_1\left(Qe_a-qK_a\right)&=&-4qe^c\frac{\delta m_c}{\delta\theta^a}\;.\label{feqmz1b}
\end{eqnarray}
Combination of the first and second equations of the system \eqref{feqmz1b} gives
\begin{eqnarray}
    R_{ab}&=&-\frac{1}{z_1}q\left(\frac{\delta m_b}{\delta e^a}-\frac{\delta m_a}{\delta e^b}\right)\;,\nonumber\\
    S_a&=&aqe_a-\frac{4}{z_1}m_a-2\frac{\kappa}{z_1^2}q\epsilon_a^{\phantom{a}bc}\left(\frac{\delta m_c}{\delta e^b}-\frac{\delta m_b}{\delta e^c}\right)\;,\label{RS1}
\end{eqnarray}
with $a$ being identified with the same $a=2\Lambda$ of \eqref{vac6} and \eqref{xi2}. Thus, the solution \eqref{RS1} reduces to the vacuum solution \eqref{vac6} as $m^a\longrightarrow0$ outside the matter distribution.

Third and fourth equations in the system \eqref{feqmz1b} can be solved, for instance, by
\begin{eqnarray}
    K_a&=&\frac{1}{2}q\theta_a-\frac{2}{z_1}q\left(\frac{\delta m^b}{\delta\omega^{ab}}+2\frac{\kappa}{z_1}\epsilon_{ab}^{\phantom{ab}c}\frac{\delta m^b}{\delta\theta^c}\right)\;,\nonumber\\
    Q&=&\frac{4}{3z_1}q\frac{\delta m^a}{\delta \theta^a}\;.\label{KQ1}
\end{eqnarray}
 The immediate conclusion is that, for the proposed solution, torsionfree solutions $Q=0$ (which allows to work with Newtonian absolute time) can only be defined in the presence of a matter distribution if the 1-form $\delta m^a/\delta\theta^b$ is traceless--remerber that $m^a$ does not depend on $q$. Obviously, a matter content that does not depend on the boost connection also ensures torsionfree solutions in the presence of matter. A third, yet improbable, possibility is that $\delta m^a/\delta\theta^a$ is parallel to $q$. We remark that this does not contradicts the previous result in Section \ref{NAT}. In fact, the result says that we can always set $Q=0$ if $K$ carries a $q$-independent sector or if the matter distribution does not depend on the boost connection. In solution \eqref{KQ1}, the reduced torsion does not have a $q$-independent part. Moreover, torsions \eqref{KQ1} reduce to torsions in \eqref{vac6} as $m^a\longrightarrow0$ outside the matter distribution.

It would be nice, however, to obtain a consistent geometry in the presence of matter together with the possibility of choosing Newtonian absolute time, \emph{i.e.}, torsionfree geometry. Nevertheless, the form of the matter action \eqref{matter0} prohibits such solution unless it does not depend on the boost connection. To show this impossibility, we maintain curvatures \eqref{RS1} and set $Q=0$ to the field equations. Thence, third and fourth equations in \eqref{feqmz1b} read now
\begin{eqnarray}
qK_a&=&\frac{4}{z_1}qe^c\frac{\delta m_c}{\delta\theta^a}\;,\nonumber\\
e_b\left(K_a-\frac{1}{2}q\theta_a\right)-e_a\left(K_b-\frac{1}{2}q\theta_b\right)&=&\frac{4}{z_1}q\left[e^c\frac{\delta m_c}{\delta\omega^{ab}}+2\frac{\kappa}{z_1}\epsilon_{abc}e^d\frac{\delta m_d}{\delta\theta_c}\right]\;.\label{KQ0b}
\end{eqnarray}
We found earlier that for a torsionfree solution be acceptable either $S_m$ must not depend on the boost connection or $K^a$ must have a piece which is independent of $q$. The first of \eqref{KQ0b} suggests that this independent part could be
\begin{equation}
    X_a=\frac{4}{z_1}e^c\frac{\delta m_c}{\delta\theta^a}\;.\label{KQ0c}
\end{equation}
Hence, the solution for $K^a$ can be obtained by combining the first of \eqref{KQ1} with \eqref{KQ0c},
\begin{equation}
    K_a=\frac{1}{2}q\theta_a-\frac{2}{z_1}q\left(\frac{\delta m^b}{\delta\omega^{ab}}+2\frac{\kappa}{z_1}\epsilon_{ab}^{\phantom{ab}c}\frac{\delta m^b}{\delta\theta^c}\right)+\frac{4}{z_1}e^c\frac{\delta m_c}{\delta\theta^a}\;.
\end{equation}
Substitution of this reduced torsion in the second of \eqref{KQ0b} results in the condition
\begin{equation}
    q\left(e_b\frac{\delta m_c}{\delta\theta^a}-e_a\frac{\delta m_c}{\delta\theta^b}\right)=0\;,\label{const3}
\end{equation}
which suggests again that the only way to accommodate Newtonian absolute time in the presence of matter for a matter content in the form \eqref{matter0} is indeed to impose that $m_c$ does not depend on the boost connection. It is also possible to show that expression \eqref{const3} is satisfied for $\delta m^a/\delta\theta^a=0$.

\section{Conclusions}\label{FINAL}

In this work we have studied the non-relativistic limit of gravity theories in the first order formalism. Particularly, the Einstein-Hilbert and Mardones-Zanelli actions were considered \cite{Mardones:1990qc,Zanelli:2005sa}. The differential form language was extensively employed. This formalism allowed us to formally obtain some interesting new results. We remark that our main strategy to solve the field equations was to look for solutions for the curvatures and torsions 2-forms. This strategy is very useful because the field equations are algebraic with respect to curvatures and torsions.

In the case of the EH action, the non-relativistic limit is widely known, see for instance \cite{Bergshoeff:2017btm,Bergshoeff:2019ctr,Bergshoeff:2017dqq}. Nevertheless, we were able to obtain some novel results which are listed below:

\begin{itemize}
    \item Starting from the field equations \eqref{feqG1} in vacuum, we have firstly obtained a solution constraining the space and time vierbeins, see \eqref{eqn}. Even though this proposal solves the field equations \eqref{Q1}, it is inconsistent with the degenerate Galilean metric nature;
    
    \item Next, in the torsionfree context, characterized by equation \eqref{Q0}, non-trivial space curvature $R$ and reduced torsion $K$ were found, see equations \eqref{R0} and \eqref{k1}. In agreement with the known literature \cite{Bergshoeff:2017dqq}. Of course, in all cases boost curvature $S^a$ is free, since the boost connection $\theta$ does not appear in the action;
    
    \item We have considered the possibility of non-trivial torsional solutions in vacuum by defining a 1-form field $n$ and the solutions \eqref{Qq1} and \eqref{Kne1}. By setting $R=S=0$, the field equations are satisfied and a kind of Newtonian Weitzenb\"ock spacetime is obtained. In this context, by setting $n$ to be constant, an explicit solution was found for the time vierbein $q$ as displayed in expression \eqref{nconst1}. A non-Newtonian time was then derived as a function of the Newtonian absolute time, see \eqref{T0}. Asymptotically, the non-Newtonian and Newtonian time coincide. Nevertheless, the obtained solution still describes a torsionfree situation since the time vierbein is parallel to $n$. In this example, by setting the spin connection to vanish, we were able to find a complete solution in terms of the explicit expressions of the vierbeins, see \eqref{vier1}. Interestingly, the corresponding spacetime metric tensors, displayed in expressions \eqref{metric3a} and \eqref{metric3b}, are conformally flat. To trivialize these metrics while respecting the conformal symmetry \eqref{conf1}, one needs to set Newtonian time by imposing $\mathsf{a}=0$. Otherwise one can only trivialize space or time metrics, but not both simultaneously;
    
    \item Finally, a solution in the presence of matter was developed under torsionfree scenario. The accidental scale symmetry \eqref{conf1} allowed us to write down the matter action in the quite general form \eqref{matter0}. Space curvature and reduced torsion were found to be directly given by the matter distribution, see \eqref{sol1} and \eqref{sol2a}. For vanishing reduced torsion one needs a matter content not depending on the spin connection. Moreover, a consistent choice constraining the space vierbein and boost connection was developed. The resulting boost curvature is given in \eqref{sol3} and \eqref{sol3a} with $S\propto K$. In this solution for the boost curvature, the vanishing case is also contemplated.
\end{itemize}  

The main novelty of the present work is the non-relativistic limit of the MZ action \eqref{mzaction1} which is the action \eqref{galaction3}. The resulting field equations are displayed in \eqref{feqmz2a}. The main results are listed below.
\begin{itemize}
    \item The first important result is that the boost-connection survives the limit and appears in the action \eqref{galaction3} and field equations \eqref{feqmz2a}. hence, the system of equations is fully determined;
    
    \item It was shown that Newtonian absolute time is well accepted by the non-relativistic limit of MZ theory, \emph{i.e.}, by the GC action \eqref{galaction3} and its field equations \eqref{feqmz2a}. By imposing torsionfree condition to the field equations we found that Newtonian absolute time can be implemented in vacuum if $qK=0$ (see equation \eqref{q00}). In the presence of matter, $K$ must have a part independent of $q$ (in this case, space torsion will be non-trivial, $T\ne0$) or the matter content must not depend on the boost connection (in this case, a torsionless solution is allowed). The typical matter field actions are \cite{Itzykson:1980rh,Weinberg:2000cr}: Klein-Gordon scalar theory; Dirac spinorial theory; Maxwell theory; Yang-Mills theories; and all their super-symmetric extensions. At least in most of them, the boost connection decouples at leading order \cite{Bergshoeff:2017btm,Hansen:2020pqs} and $\sigma^a=0$ would be a safe choice;
    
    \item We have considered a generic external matter distribution of the form \eqref{matter0} in the context of GC gravity. In such situation $m_a$ may depend on all fields, except $q$. We have found the curvatures \eqref{RS1} and torsions \eqref{KQ1}. These configurations reduce to the vacuum solutions \eqref{vac6} in the limit $m_a\longrightarrow0$. Particularly, the second of \eqref{KQ1} says that, for the matter content of the form \eqref{matter0}, Newtonian absolute time can be employed in the presence of matter only if at least one of the following conditions is satisfied: $m_c$ do not depend on the boost connection; the 1-form $\delta m^a/\delta\theta^b$ is parallel to $q$; the 1-form $\delta m^a/\delta\theta^b=0$ is traceless.
    
    \item Finally, in the Appendix \eqref{CCap}, it was shown that if one enforces the cosmological constant term to survive the limit, the resulting theory is inconsistent: The vacuum solution does not agree with the non-relativistic limit of the maximally symmetric solution \eqref{dS0}; The vacuum solution is inconsistent with the Bianchi identities \eqref{Bia4}. This result shows that the vanishing of the cosmological constant in the leading order of the non-relativistic limit at leading order is inevitable.
    
\end{itemize}
 
Although many physical insights were obtained from our formal analysis, it remains to apply the results to actual physical systems. These tests could provide some interesting bounds for some parameters of the MZ action, specifically $z_1$. Moreover, new effects could emerge from these generalized gravity theory which could be tested in the non-relativistic context. Another point to be explored is the fact that MZ theorem holds for any spacetime dimension. So, generalization of the present results in other dimensions could be developed in the future. Furthermore, one could consider the usual extensions of the Galilei group, such as Bargmann and Schr\"odinger groups, in order to construct more general theories \cite{Bergshoeff:2019ctr}. As mentioned in the Introduction, these theories could be relevant in Ho$\check{\mathrm{r}}$ava-Lifshits  \cite{Afshar:2015aku} and string \cite{Bergshoeff:2019pij,Bergshoeff:2020oes} contexts.

\section*{Acknowledgements}

The authors are in debt with R.~de Melo e Souza, A.~D.~Pereira, G.~Sadovski, and A.~A.~Tomaz for the critical review of the manuscript and enlightening discussions. This study was financed in part by The Coordena\c c\~ao de Aperfei\c coamento de Pessoal de N\'ivel Superior - Brasil (CAPES) - Finance Code 001. 
\appendix

\section{The fate of the cosmological constant}\label{CCap}

In this appendix we discuss what happens if one chooses not to scale the cosmological constant, \emph{i.e.}, $\Lambda\longrightarrow\Lambda$ instead of $\Lambda\longrightarrow\zeta^{-2}\Lambda$ as in \eqref{resc3}. This choice would be motivated if one chooses to rescale the parameters only to avoid possible divergences in the limit $\zeta\longrightarrow\infty$ (See footnote 10). In this case, action \eqref{galaction1} reduces to
\begin{equation}
  S_{GMZ}=2\kappa\int\;\epsilon_{abc}q\left(R^{ab}e^c+\Lambda e^ae^be^c\right)+\int\left[z_1\left(R^{ab}e_ae_b-qe_aS^a\right)+z_2Q^2+z_4R^{ab}R_{ab}\right]+S_m\;.\label{galaction2}
\end{equation}
in its non-relativistic limit. The corresponding field equations read now
\begin{eqnarray}
2\kappa\epsilon_{abc}\left(R^{ab}e^c+\Lambda e^ae^be^c\right)-z_1e_aS^a&=&-\tau\;,\nonumber\\
2\kappa\epsilon_{abc}q\left(R^{bc}+3\Lambda e^be^c\right)-z_1\left(2R_{ab}e^b+qS_a\right)&=&\tau_a\;,\nonumber\\
2\kappa D(\epsilon_{abc}qe^c)+z_1D(e_ae_b)-\frac{z_1}{2}q(e_a\theta_b-e_b\theta_a)&=&-\sigma_{ab}\;,\nonumber\\
z_1\left(Qe^a-qK^a\right)&=&\sigma_a\;.\label{feqmz1}
\end{eqnarray}
Combination of the first with the second equation of the system \eqref{feqmz1} imply on
\begin{equation}
    4\kappa\Lambda \epsilon_{abc}qe^ae^be^c+2z_1R^{ab}e_ae_b=q\tau-e^a\tau_a\;.\label{comb1a}
\end{equation}
Combining the third and fourth equations of the system \eqref{feqmz1} results in
\begin{equation}
e_b\left(K_a-\frac{1}{2}q\theta_a\right)-e_a\left(K_b-\frac{1}{2}q\theta_b\right)=-\frac{1}{z_1}\sigma_{ab}-2\frac{\kappa}{z^2_1}\epsilon_{abc}\sigma^c\;.\label{comb1b}
\end{equation}

The field equations \eqref{feqmz1} in vacuum read
\begin{eqnarray}
2\kappa\epsilon_{abc}\left(R^{ab}e^c+\Lambda e^ae^be^c\right)-z_1e_aS^a&=&0\;,\nonumber\\
2\kappa\epsilon_{abc}q\left(R^{bc}+3\Lambda e^be^c\right)-z_1\left(2R_{ab}e^b+qS_a\right)&=&0\;,\nonumber\\
2\kappa D(\epsilon_{abc}qe^c)+z_1D(e_ae_b)-\frac{z_1}{2}q(e_a\theta_b-e_b\theta_a)&=&0\;,\nonumber\\
Qe^a-qK^a&=&0\;.\label{feqmz1vac}
\end{eqnarray}
It can be straightforwardly checked that the first two equations in \eqref{feqmz1vac} accepts a general solution with curvatures of the form
\begin{eqnarray}
    R^{ab}&=&-2\frac{\kappa\Lambda}{z_1}\epsilon^{ab}_{\phantom{ab}c}qe^c+(b\Lambda+cz_1) e^ae^b\;,\nonumber\\
    S^a&=&2\kappa\left[\frac{\Lambda}{z_1}(1+b)+c\right]\epsilon^a_{\phantom{a}bc}e^be^c+aqe^a\;,\label{vac5}
\end{eqnarray}
with $b$ and $c$ being dimensionless parameters and $a$ a mass squared parameter. Note that, due to the presence of the cosmological constant, vanishing curvatures do not satisfy the first two field equations \eqref{feqmz1vac}. Substitution of \eqref{vac5} in the third Bianchi identity in \eqref{Bia2} implies on the relation 
\begin{equation}
    4\kappa\left[\frac{\Lambda}{z_1}(1+b)+c\right]\epsilon^a_{\phantom{a}bc}K^be^c=2\frac{\kappa\Lambda}{z_1}\epsilon^a_{\phantom{a}bc}q\theta^be^c+(b\Lambda+cz_1) e^ae_b\theta^b\;,\label{rel2a}
\end{equation}
where the fourth equation in \eqref{feqmz1vac} was employed. On the other hand, curvatures \eqref{vac5} relate to each other through
\begin{equation}
    \frac{\Lambda}{z_1}qS^a=\left[\frac{\Lambda}{z_1}(1+b)+c\right]R^a_{\phantom{a}b}e^b\;.\label{rel2b}
\end{equation}
Applying the covariant derivative in relation \eqref{rel2b} one finds
\begin{equation}
     \frac{\Lambda}{z_1}QS^a- \frac{\Lambda}{z_1}qR^a_{\phantom{a}b}\theta^b=\left[\frac{\Lambda}{z_1}(1+b)+c\right]R^a_{\phantom{a}b}K^b\;.\label{rel2c}
\end{equation}
Now, for the theory to be physically consistent, a torsionless vacuum solution should be acceptable,
\begin{eqnarray}
    Q&=&0\;,\nonumber\\
    K^a&=&\frac{1}{2}q\theta^a\;.\label{torsionless1}
\end{eqnarray}
This solution satisfies the third and fourth field equations in \eqref{feqmz1}. For this solution be consistent with the relation \eqref{rel2a}, one can quickly verify that we need to set
\begin{equation}
    c=-\frac{\Lambda}{z_1}b\;.\label{consts1}
\end{equation}
However, \eqref{torsionless1} together with \eqref{consts1} do not satisfy \eqref{rel2c}. In fact, \eqref{rel2c} will only be satisfied if the cosmological constant is set to zero. This is important if one wishes torsionless vacuum solutions together with non-trivial curvatures which might be associated with the mass parameters of the theory. The present analysis suggests that is safer to consider that the cosmological constant must indeed scale as $\Lambda\longrightarrow\zeta^{-\xi}\Lambda$ with $\xi$ being some dimensionless positive real number to be determined. In the present work we followed the know literature \cite{Aldrovandi:1998im,Hansen:2020pqs} and employed $\xi=2$.

\bibliographystyle{utphys2}
\bibliography{library}

\providecommand{\href}[2]{#2}\begingroup\raggedright\begin{thebibliography}{10}

\bibitem{Misner:1974qy}
C.~W. Misner, K.~Thorne, and J.~Wheeler, {\em {Gravitation}}.
\newblock W. H. Freeman, San Francisco, 1973.

\bibitem{Wald:1984rg}
R.~M. Wald,
  \href{http://dx.doi.org/10.7208/chicago/9780226870373.001.0001}{{\em {General
  Relativity}}}.
\newblock Chicago Univ. Pr., Chicago, USA, 1984.

\bibitem{DeSabbata:1986sv}
V.~De~Sabbata and M.~Gasperini, {\em {Introduction to Gravity}}.
\newblock World Scientific, 346p, Singapore, 1986.

\bibitem{Utiyama:1956sy}
R.~Utiyama, ``{Invariant theoretical interpretation of interaction}''.
\href{http://dx.doi.org/10.1103/PhysRev.101.1597}{{\em Phys. Rev.} {\bfseries
  101} (1956) 1597--1607}.

\bibitem{Kibble:1961ba}
T.~W.~B. Kibble, ``{Lorentz invariance and the gravitational field}''.
\href{http://dx.doi.org/10.1063/1.1703702}{{\em J. Math. Phys.} {\bfseries 2}
  (1961) 212--221}.

\bibitem{Sciama:1964wt}
D.~W. Sciama, ``{The Physical structure of general relativity}''.
  \href{http://dx.doi.org/10.1103/RevModPhys.36.1103}{{\em Rev. Mod. Phys.}
  {\bfseries 36} (1964) 463--469}.
[Erratum: Rev. Mod. Phys.36, 1103(1964)].

\bibitem{Mardones:1990qc}
A.~Mardones and J.~Zanelli, ``{Lovelock-Cartan theory of gravity}''.
\href{http://dx.doi.org/10.1088/0264-9381/8/8/018}{{\em Class. Quant. Grav.}
  {\bfseries 8} (1991) 1545--1558}.

\bibitem{Zanelli:2005sa}
J.~Zanelli, ``{Lecture notes on Chern-Simons (super-)gravities. Second edition
  (February 2008)}''. in {\em {Proceedings, 7th Mexican Workshop on Particles
  and Fields (MWPF 1999): Merida, Mexico, November 10-17, 1999}}.
\newblock
2005.
\newblock

\bibitem{Inonu:1953sp}
E.~Inonu and E.~P. Wigner, ``{On the Contraction of groups and their
  represenations}''. \href{http://dx.doi.org/10.1073/pnas.39.6.510}{{\em Proc.
  Nat. Acad. Sci.} {\bfseries 39} (1953) 510--524}.

\bibitem{Bergshoeff:2017btm}
E.~Bergshoeff, J.~Gomis, B.~Rollier, J.~Rosseel, and T.~ter Veldhuis,
  ``{Carroll versus Galilei Gravity}''.
\href{http://dx.doi.org/10.1007/JHEP03(2017)165}{{\em JHEP} {\bfseries 03}
  (2017) 165}.

\bibitem{Banerjee:2018gqz}
R.~Banerjee and P.~Mukherjee, ``{Galilean gauge theory from Poincare gauge
  theory}''. \href{http://dx.doi.org/10.1103/PhysRevD.98.124021}{{\em Phys.
  Rev. D} {\bfseries 98} no.~12, (2018) 124021}.

\bibitem{Christensen:2013rfa}
M.~H. Christensen, J.~Hartong, N.~A. Obers, and B.~Rollier, ``{Boundary
  Stress-Energy Tensor and Newton-Cartan Geometry in Lifshitz Holography}''.
  \href{http://dx.doi.org/10.1007/JHEP01(2014)057}{{\em JHEP} {\bfseries 01}
  (2014) 057}.

\bibitem{Banerjee:2014nja}
R.~Banerjee, A.~Mitra, and P.~Mukherjee, ``{Localization of the Galilean
  symmetry and dynamical realization of Newton-Cartan geometry}''.
  \href{http://dx.doi.org/10.1088/0264-9381/32/4/045010}{{\em Class. Quant.
  Grav.} {\bfseries 32} no.~4, (2015) 045010}.

\bibitem{Afshar:2015aku}
H.~R. Afshar, E.~A. Bergshoeff, A.~Mehra, P.~Parekh, and B.~Rollier, ``{A
  Schrödinger approach to Newton-Cartan and Ho\v rava-Lifshitz gravities}''.
  \href{http://dx.doi.org/10.1007/JHEP04(2016)145}{{\em JHEP} {\bfseries 04}
  (2016) 145}.

\bibitem{Abedini:2019voz}
M.~Abedini, H.~R. Afshar, and A.~Ghodsi, ``{Covariant Poisson's equation in
  torsional Newton-Cartan gravity}''.
  \href{http://dx.doi.org/10.1007/JHEP04(2019)117}{{\em JHEP} {\bfseries 04}
  (2019) 117}.

\bibitem{Banerjee:2016laq}
R.~Banerjee and P.~Mukherjee, ``{Torsional Newton\textendash{}Cartan geometry
  from Galilean gauge theory}''.
  \href{http://dx.doi.org/10.1088/0264-9381/33/22/225013}{{\em Class. Quant.
  Grav.} {\bfseries 33} no.~22, (2016) 225013}.

\bibitem{Bergshoeff:2017dqq}
E.~Bergshoeff, A.~Chatzistavrakidis, L.~Romano, and J.~Rosseel,
  ``{Newton-Cartan Gravity and Torsion}''.
\href{http://dx.doi.org/10.1007/JHEP10(2017)194}{{\em JHEP} {\bfseries 10}
  (2017) 194}.

\bibitem{Cartan:1923zea}
E.~Cartan, ``{Sur les vari\'et\'es \`a connexion affine et la th\'eorie de la
  relativit\'e g\'en\'eralis\'ee. (premi\`ere partie)}''. {\em Annales Sci.
  Ecole Norm. Sup.} {\bfseries 40} (1923) 325--412.

\bibitem{Cartan:1924yea}
E.~Cartan, ``{Sur les vari\'et\'es \`a connexion affine et la th\'eorie de la
  relativit\'e g\'en\'eralis\'ee. (premi\`ere partie) (Suite).}''. {\em Annales
  Sci. Ecole Norm. Sup.} {\bfseries 41} (1924) 1--25.

\bibitem{Trautman:1963aaa}
A.~Trautman, ``{Sur la th\'eorie newtonienne de la gravitation}''. {\em C. R.
  Acad. Sci. Paris} {\bfseries 257} (1963) 617--620.

\bibitem{Havas:1964zza}
P.~Havas, ``{Four-Dimensional Formulations of Newtonian Mechanics and Their
  Relation to the Special and the General Theory of Relativity}''.
  \href{http://dx.doi.org/10.1103/RevModPhys.36.938}{{\em Rev. Mod. Phys.}
  {\bfseries 36} (1964) 938--965}.

\bibitem{Trautman:1965aaa}
A.~Trautman, {\em {Lectures on general relativity}}.
\newblock Engtewood Cliffs, N. J.: Prentice-Hall, Brandeis Summer Institute,
  1965.

\bibitem{Kunzle:1972aaa}
H.~P. Künzle, ``{Galilei and Lorentz structures on space-time: comparison of
  the corresponding geometry and physics}''. {\em Ann. Inst. Henri Poincar\'e}
  {\bfseries 17} (1972) 337---362.

\bibitem{Dixon:1975fy}
W.~Dixon, ``{On the Uniqueness of the Newtonian Theory as a Geometric Theory of
  Gravitation}''. \href{http://dx.doi.org/10.1007/BF01629247}{{\em Commun.
  Math. Phys.} {\bfseries 45} (1975) 167--182}.

\bibitem{Duval:1984cj}
C.~Duval, G.~Burdet, H.~Kunzle, and M.~Perrin, ``{Bargmann Structures and
  Newton-cartan Theory}''.
  \href{http://dx.doi.org/10.1103/PhysRevD.31.1841}{{\em Phys. Rev. D}
  {\bfseries 31} (1985) 1841--1853}.

\bibitem{Duval:2009vt}
C.~Duval and P.~A. Horvathy, ``{Non-relativistic conformal symmetries and
  Newton-Cartan structures}''.
  \href{http://dx.doi.org/10.1088/1751-8113/42/46/465206}{{\em J. Phys. A}
  {\bfseries 42} (2009) 465206}.

\bibitem{Bergshoeff:2019ctr}
E.~Bergshoeff, J.~M. Izquierdo, T.~Ortín, and L.~Romano, ``{Lie Algebra
  Expansions and Actions for Non-Relativistic Gravity}''.
\href{http://dx.doi.org/10.1007/JHEP08(2019)048}{{\em JHEP} {\bfseries 08}
  (2019) 048}.

\bibitem{Hartong:2015zia}
J.~Hartong and N.~A. Obers, ``{Ho\v{r}ava-Lifshitz gravity from dynamical
  Newton-Cartan geometry}''.
  \href{http://dx.doi.org/10.1007/JHEP07(2015)155}{{\em JHEP} {\bfseries 07}
  (2015) 155}.

\bibitem{Horava:2009uw}
P.~Horava, ``{Quantum Gravity at a Lifshitz Point}''.
  \href{http://dx.doi.org/10.1103/PhysRevD.79.084008}{{\em Phys. Rev. D}
  {\bfseries 79} (2009) 084008}.

\bibitem{Roychowdhury:2019sfo}
D.~Roychowdhury, ``{Semiclassical dynamics for torsional Newton-Cartan
  strings}''. \href{http://dx.doi.org/10.1016/j.nuclphysb.2020.115132}{{\em
  Nucl. Phys. B} {\bfseries 958} (2020) 115132}.

\bibitem{Kluson:2020aoq}
J.~Kluso\v{n}, ``{Unstable D-brane in Torsional Newton-Cartan Background}''.
  \href{http://dx.doi.org/10.1007/JHEP09(2020)191}{{\em JHEP} {\bfseries 09}
  (2020) 191}.

\bibitem{Bergshoeff:2019pij}
E.~A. Bergshoeff, J.~Gomis, J.~Rosseel, C.~\c{S}im\c{s}ek, and Z.~Yan,
  ``{String Theory and String Newton-Cartan Geometry}''.
  \href{http://dx.doi.org/10.1088/1751-8121/ab56e9}{{\em J. Phys. A} {\bfseries
  53} no.~1, (2020) 014001}.

\bibitem{Bergshoeff:2020oes}
E.~Bergshoeff, J.~Lahnsteiner, L.~Romano, and C.~Simsek, ``{Non-relativistic
  String theory}''. \href{http://dx.doi.org/10.22323/1.376.0146}{{\em PoS}
  {\bfseries CORFU2019} (2020) 146}.

\bibitem{Lovelock:1971yv}
D.~Lovelock, ``{The Einstein tensor and its generalizations}''.
\href{http://dx.doi.org/10.1063/1.1665613}{{\em J. Math. Phys.} {\bfseries 12}
  (1971) 498--501}.

\bibitem{Weitzenbock:1923boo}
R.~Weitzenbock, {\em {Invariantentheorie}}.
\newblock Noordhoff, Groeningen,
1923.
\newblock

\bibitem{Garecki:2010jj}
J.~Garecki, ``{Teleparallel equivalent of general relativity: A Critical
  review}''. in {\em {Hypercomplex Seminar 2010: (Hyper)Complex and
  Randers-Ingarden Structures in Mathematics and Physics}}.
\newblock 10, 2010.

\bibitem{Aldrovandi:2013wha}
R.~Aldrovandi and J.~G. Pereira,
  \href{http://dx.doi.org/10.1007/978-94-007-5143-9}{{\em {Teleparallel
  Gravity}: {An Introduction}}}, vol.~173.
\newblock Springer, 2013.

\bibitem{Golovnev:2018red}
A.~Golovnev, ``{Introduction to teleparallel gravities}''. in {\em {9th
  Mathematical Physics Meeting}: {Summer School and Conference on Modern
  Mathematical Physics}}.
\newblock 1, 2018.

\bibitem{Nieh:1981ww}
H.~Nieh and M.~Yan, ``{An Identity in Riemann-cartan Geometry}''.
  \href{http://dx.doi.org/10.1063/1.525379}{{\em J. Math. Phys.} {\bfseries 23}
  (1982) 373}.

\bibitem{Nieh:2007zz}
H.~T. Nieh, ``{A torsional topological invariant}''.
\href{http://dx.doi.org/10.1142/S0217751X07038414}{{\em Int. J. Mod. Phys.}
  {\bfseries A22} (2007) 5237--5244}.

\bibitem{Nieh:2018rlg}
H.~Nieh, ``{Torsional Topological Invariants}''.
  \href{http://dx.doi.org/10.1103/PhysRevD.98.104045}{{\em Phys. Rev. D}
  {\bfseries 98} no.~10, (2018) 104045}.

\bibitem{Kobayashi}
S.~Kobayashi and K.~Nomizu, {\em {Foundations of Differential Geometry}}.
\newblock John Wiley \& Sons, Inc. New York, US. 329p,
1963.
\newblock

\bibitem{Nakahara:1990th}
M.~Nakahara, {\em {Geometry, topology and physics}}.
\newblock Taylor \& Francis, Bristol, UK: Hilger 505 p. (Graduate student
  series in physics),
1990.
\newblock

\bibitem{Hansen:2018ofj}
D.~Hansen, J.~Hartong, and N.~A. Obers, ``{Action Principle for Newtonian
  Gravity}''. \href{http://dx.doi.org/10.1103/PhysRevLett.122.061106}{{\em
  Phys. Rev. Lett.} {\bfseries 122} no.~6, (2019) 061106}.

\bibitem{Aldrovandi:1998im}
R.~Aldrovandi, A.~Barbosa, L.~Crispino, and J.~Pereira, ``{Non-Relativistic
  spacetimes with cosmological constant}''.
  \href{http://dx.doi.org/10.1088/0264-9381/16/2/013}{{\em Class. Quant. Grav.}
  {\bfseries 16} (1999) 495--506}.

\bibitem{Hansen:2020pqs}
D.~Hansen, J.~Hartong, and N.~A. Obers, ``{Non-Relativistic Gravity and its
  Coupling to Matter}''. \href{http://dx.doi.org/10.1007/JHEP06(2020)145}{{\em
  JHEP} {\bfseries 06} (2020) 145}.

\bibitem{Itzykson:1980rh}
C.~Itzykson and J.~B. Zuber, {\em {Quantum Field Theory}}.
\newblock International Series In Pure and Applied Physics. McGraw-Hill, New
  York, 1980.
\newblock
\url{http://dx.doi.org/10.1063/1.2916419}.
\newblock

\bibitem{Weinberg:2000cr}
S.~Weinberg, {\em {The quantum theory of fields. Vol. 3: Supersymmetry}}.
\newblock Cambridge University Press, 6, 2013.

\end{thebibliography}\endgroup

\end{document}